\documentclass[prl,twocolumn,superscriptaddress,showpacs]{revtex4}
\usepackage{color,psfig,graphics,graphicx,psfrag,amsmath,amssymb,amsfonts,rotating}

\begin{document}

\title{Lossy data compression with random gates}

\author{Stefano Ciliberti}

\affiliation{Laboratoire de Physique Th\'eorique et Mod\`eles Statistiques,
Universit\'e de Paris-Sud, b\^atiment 100, 91405, Orsay Cedex, France}

\author{Marc M\'ezard}

\affiliation{Laboratoire de Physique Th\'eorique et Mod\`eles Statistiques,
Universit\'e de Paris-Sud, b\^atiment 100, 91405, Orsay Cedex, France}

\author{Riccardo Zecchina}

\affiliation{ICTP, Strada Costiera 11, I-34100 Trieste, Italy}

\date{\today}

\begin{abstract}
We introduce a new protocol for a lossy data compression algorithm which is
based on constraint satisfaction gates. We show that the theoretical
capacity of algorithms built from standard parity-check gates converges
exponentially fast to the Shannon's bound when the number of variables seen
by each gate increases. We then generalize this approach by introducing
random gates. They have theoretical performances nearly as good as parity
checks, but they offer the great advantage that the encoding can be done in
linear time using the Survey Inspired Decimation algorithm, a powerful
algorithm for constraint satisfaction problems derived from statistical
physics.
\end{abstract}


\maketitle

\emph{Introduction.} Constraint satisfaction problems (CSPs) are at the
heart of an emerging field of research which is of interest to statistical
physics, combinatorial optimization, statistical inference and information
theory~\cite{review}. Broadly speaking, these are problems involving a
large number of variables, taking values in a finite set (hereafter we shall
keep to binary variables). Each constraint involves $K$ variables, and
imposes a probability law on the $2^K$ possible assignments of the
variables in this subset. Hard constraints just forbid some of the
configurations. The spin glass problem~\cite{spinglass}, the satisfiability problem
which lies at the heart of the theory of computational complexity in
computer science~\cite{papa}, or the parity check problems used in error
correcting codes~\cite{ecc} all belong to this category.

A lot of progress has been made in recent years in the study of random
constraint satisfaction problems where each constraint involves randomly
chosen (with uniform distribution) variables~\cite{SPgeneral}. This is the
natural setting for spin glasses, it offers the possibility to study issues
in typical case complexity in satisfiability, and it provides some of the
most efficient codes for error correction. In several cases, it has been
found that when the density of constraints increases the system enters first
a 'clustered' phase before it reaches the threshold of unsatisfiability
where it cannot meet all the constraints. Above this threshold the
configurations which violate the smaller number of constraints are also
clustered. Clustering means that the configurations which satisfy all the
constraints are grouped into many disconnected clusters which are distant
from each other. Statistical physics methods originating from spin glass
theory, like the replica and cavity methods, turn out to be very efficient
to study these phenomena~\cite{mozeki,meze}, and some of the results have
been confirmed rigorously
recently~\cite{merize,montanariricci,cocco,mora}. They have also led to a
powerful algorithm (survey propagation) which is able to solve very large
problems in the clustered region~\cite{meze}.

We will show how one can take advantage of these clustered phases to address
a classic problem in coding theory, lossy data compression. While a large
amount of work has been done in this field~\cite{bergergibson}, a number of
challenging problems are open, among them the realization of a practical
compression protocol for correlated sources or the exponential increasing
time in the encoding/decoding step of typical algorithms. As for lossy
compression schemes, it is worth to mention in particular the good
performance of algorithms based on the codes~\cite{caire} developed in the
context of channel coding. Here we propose an alternative strategy, and as a
starting point we focus on the case of uncorrelated sources.

The problem of lossy data compression can be summarized as follows. We have
a source alphabet ${\cal A}$, a source distribution $p(x_a)$ and a
distortion measure $d(\cdot,\cdot)$ which takes values in $[0,1]$. We start
from an original message which is a sequence $\{ x_a\}$ of $M$ values
independently drawn from the source distribution. The purpose of data
compression is to map this message to a string of $N$ bits, with $N<M$, in
such a way that they can be for example easily transmitted or stored. Then,
one wants to decode this $N$-bits string in order to reconstruct a sequence
as close as possible to the original message. We call the decoded message
$\{ x_a^*\}$ and we want to minimize the expected value of the distortion $D
\equiv {\mathbb E}\sum_{a=1}^M d(x_a,x_a^*)/M$. How small a distortion one
can achieve depends on the rate $R=N/M$ at which the original message has
been compressed.

The rate distortion theorem proved by Shannon~\cite{shannonlossy} provides a
bound for the minimum rate at which a compression is possible once we fix
the average distortion we tolerate. The analytic expression of this
\emph{rate-distortion function} $R(D)$ is not known explicitly in the most
general case of correlated (memory) sources, and it is most often obtained
by means of some numerical algorithm (see {\it e.g.}~\cite{blahut}). On the
other hand, for an uncorrelated unbiased binary source ({\it i.e.}  with
$p(x_a=0)=1-p(x_a=1)=1/2$), the rate-distortion function in the large $N$
limit has the simple expression
\begin{equation}
  R(D) = 1 + D\log_2 D +(1-D)\log_2(1-D)\ .
  \label{eq:shannon}
\end{equation}
As a first requirement, a good lossy data compression algorithm must be able
to approach this theoretical limit. One should mention that in the lossless
case, that is the $D\to 0$ limit, practical algorithms that saturate
asymptotically the Shannon's bound have been discovered a long time
ago~\cite{lossless}. The lossy case turns out to be more difficult from the
algorithmic point of view. Recently, a perceptron with a non-linear transfer
function has been proposed~\cite{perceptron} as a lossy compressor and it
has been analytically shown to achieve theoretically optimal performance,
but its practical use is strongly reduced by the fact that there is no known
polynomial algorithm in this case, and the typical string lengths that can
be compressed in reasonable time are thus rather short.

\begin{figure}
  \psfrag{x=x1x2x3x4x5x6x7x8}[][][2.3]
         {$x\ =\ \{\ x_1\;\; x_2\;\; x_3\;\; x_4\;\; x_5\;\; x_6\;\; x_7\;\; x_8\; \}\;$}
  \psfrag{y=y1y2y3y4y5}[][][2.3]
         {$y\ =\ \{\;\; y_1\quad y_2\quad y_3\quad y_4\quad y_5\; \}\;\;$}
  \begin{center}
    \includegraphics[angle=0,width=0.35\textwidth]{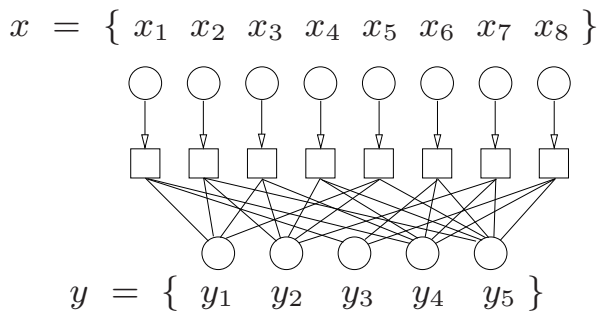}
  \end{center}
  \caption{A Tanner graph is a convenient representation for a CSP. We
  emphasize in this cartoon the topological support of our protocol.}
\label{fig:graph}
\end{figure}

\emph{The general idea.} We are interested here in developing an approach to
the problem which is based on CSPs, as suggested in~\cite{martinianyedidia}.
Our CSP uses $M$ constraints between $N$ Boolean variables taking values in
$\{0,1\}$.  Each constraint, say $a$, is actually a gate controlled by the
value $x_a$ of the $a$-th bit of the original message (see
Fig.~\ref{fig:graph}). The gate $a$ is connected to $K_a$ randomly chosen
variables. The $2^{K_a}$ possible configurations of these variables are
partitioned into two equal size subsets, $S_a$ and $U_a$.  When the control
bit is $x_a=0$, the configurations in $S_a$ satisfy the constraint, the
configurations in $U_a$ don't. When the control bit is $x_a=1$, the
configurations in $U_a$ satisfy the constraint, the configurations in $S_a$
don't. A simple example is provided by parity checks.  $S_a$ consists of the
configurations with an even number of $0$'s. The gate then performs a linear
operation: it checks whether the sum of all variables and $b_a$ is equal to
$0$ modulo 2.  In this case the CSP is nothing but the well known XORSAT
problem in computer science~\cite{xor}; it can also been seen as a spin
glass problem with three-spin interactions.

In our procedure, the initial word of $M$ bits is used to build a CSP with
$M$ constraints for $N$ $(<M)$ variables $\{y_1,\ldots y_N\}$. We then look
for a configuration of variables ${\bf y^*}$ that minimizes the number of
violated constraints, that is ground state configuration. This is the
encoded (compressed) word. Of course, this step is non-trivial since one
must be able to handle a CSP which is in its ``unsat'' phase. We note that
the rate $R$ of the process is simply related to the density of constraints
$\alpha=M/N$ by $R=1/\alpha$. Once we have a ground state configuration, the
decoding is easy: for each constraint $a$, one considers the configuration
of the subset of $K_a$ variables appearing in $a$. If it lies in $S_a$, the
reconstructed bit is $x_a^*= 0$, otherwise it is $x_a^*= 1$.  The number of
bits of the original message which are wrongly reconstructed is nothing but
the number of constraints violated in a ground state configuration.

We shall measure the distortion as $D=\sum_{a=1}^M |x_a-x^*_a|/M$.  We
define the total ``energy'' of a configuration ${\bf y}$ of the CSP as
$E({\bf y})=\sum_{a=1}^M \varepsilon_a$, where $\varepsilon_a({\bf y})=0$ if
the constraint $a$ is satisfied by the global configuration ${\bf y}$, and
$\varepsilon_a({\bf y})=1$ otherwise.  The distortion is then related to the
ground state energy $E_0$ of the CSP through:
\begin{equation}
  D=E/M\ .
  \label{eq:distortion}
\end{equation}
We are interested in the thermodynamic limit $N,M\to \infty$ at fixed
density of constraints $\alpha$. Shannon's theorem provides a lower bound
$E_{Sh}(\alpha)$ to the ground state energy, and a good compression
algorithm should be based on a CSP with a very low ground state energy, as
near as possible to Shannon's value.

The coder based on parity check gates (XORSAT-CSP) is a good candidate. A
general strategy for computing the ground state energy of this problem has
been developed in~\cite{merize}. When all checks involve $K$ variables and
$K$ becomes large, a computation based on this strategy~\cite{semprenoi}
shows that $E_0(\alpha) - E_{Sh}(\alpha)$ decreases exponentially with
$K$. So the theoretical capacity of these gates rapidly approach Shannon's
limit when $K$ increases. Unfortunately there is no known algorithm which
matches this theoretical capacity.  This is in contrast to the use of low
density parity check codes for channel coding, where message passing
techniques are known to perform quite well. However we shall see below that
message passing does perform well on some other classes of gates.

\emph{Message passing} Useful techniques for solving random CSPs are based
on local iterative updates of some ``messages'' sent along the graph. For
example, applying the `Min-Sum' algorithm~\cite{minsum} to CSPs, one obtains
the Warning Propagation (WP) algorithm: each constraint $a$ sends a warning
message $u_{a \to i}$ to one neighbor variable $i$ according to the values
of the other variables attached to it: This message can be $0$ -- meaning
that the variable is free to assume any value~\footnote{A variable is free
when the status of the constraint ($\varepsilon_a=0$ or $1$) is independent
of the actual value of the variable.}  -- or $1$ -- meaning that, in order
to satisfy that constraint, the variable should assume a certain value
--. This algorithm is very powerful and efficient in many CSPs where the
underlying factor graph is locally tree-like, when the density of
constraints is small enough.  However it is limited in two aspects:  1) it
stops to converge when the density of constraint is such that the system is
in a clustered phase, and in particular in the unsat regime which is of
interest for our compression scheme. 2) It does not work for parity checks
because of the basic symmetry of these gates.

The first limitation can be handled by going to a more sophisticated message
passing algorithm, Survey Propagation (SP). The SP algorithm is the direct
implementation of the 1RSB cavity equations on a single sample. In this
case, one works with the probability distribution $Q_{a\to i}(u_{a\to i})$
that, if we pick one cluster at random, the warning $u_{a\to i}$ is sent
along the link from constraint $a$ to variable $i$. This is the general
object which is needed in order to cope with the appearance of disconnected
clusters of solutions. Once we know the probabilities of all the warnings,
we estimate the probability of the total \emph{bias} $H_i$ on the variable
$i$ ,defined as $ H_i = \sum_{a\in V(i)} u_{a\to i}\ , $ where $V(i)$ is the
set of constraints attached to the variable $i$.  The idea behind the Survey
Inspired Decimation (SID,~\cite{meze}) is to take advantage of this
information to fix the most biased variable in the system. Once this is
done, oen has a simplified CSP with $N-1$ variables and we can thus repeat
this step until one is left with an under-constrained problem solvable by
some standard local algorithm. This algorithm has been shown to be very
useful in CSP problems like the coloring or the K-SAT, where one is able to
find efficiently a SAT assignment in the difficult region. It can also be
used for problems with higher constraint density in order to find
configurations of low energy (small number of violated
constraints)~\cite{battaglia}, which makes it a very useful tool for
compression.

\emph{Non-linear nodes.}  While it performs well on many CSPs, the SP
algorithm is useless for the ``parity source coder''. The problem here comes
from the fact that the distribution of the total bias is always symmetric,
so that the messages obtained after convergence give no hint of how to
decimate the problem.  In order to solve this problem, we propose a
compression scheme based on some other gates, different from parity
checks. These turn out to have a theoretical capacity close to the parity
checks, and a generalized version of the SP algorithm~\cite{semprenoi} leads
to a convergent decimation scheme which is an efficient coding algorithm.
Among the several types of constraints we have examined, the ``random''
nodes have been found to be the more efficient. They are defined as follows:
The subset $S_a$ is a randomly chosen subset of size $2^{K-1}$.  While
parity checks just implement linear constraints on $Z_2$, these random nodes
are non-linear functions of their inputs. However from the point of view of
the cavity method (used to compute their theoretical performance) and of the
SP message passing procedure (used as encoding algorithm), they can be
handled by relatively straightforward generalizations of the methods used
for parity checks~\cite{semprenoi}.  In this note we summarize the results.

\begin{figure}
  \begin{center}
    \includegraphics[angle=-90,width=0.5\textwidth]{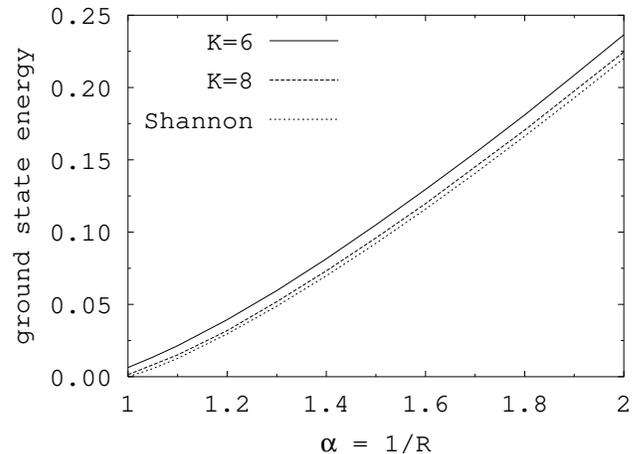}
  \end{center}
  \caption{The ground state energy of the CSP based on non-linear nodes
    versus the constraint density $\alpha$. Shannon's bound is also plotted
    for comparison.}
\label{fig:capacity}
\end{figure}
\begin{figure}
  \begin{center}
    \includegraphics[angle=-90,width=0.5\textwidth]{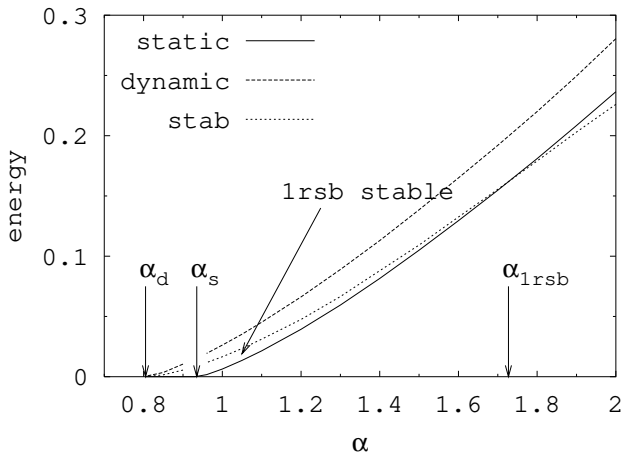}
  \end{center}
  \caption{The phase diagram of a system with $30$ different random nodes
  with $K=6$. The values for the thresholds are: $\alpha_d=0.803$,
  $\alpha_s=0.935$ and $\alpha_{1rsb}=1.727$. }
\label{fig:phasediagram}
\end{figure}

We build up a list of $r$ random checks, and each constraint $a$ picks up
one check randomly in this list~\footnote{We forbid ``fully canalizing''
nodes that just depend on one variable.}. This allows to memorize the truth
table of all nodes and thus to speed up the algorithms.  All the results
quoted below are for $r=30$. The theoretical capacity of this system, which
is proportional to the ground state energy according to
(\ref{eq:distortion}), is illustrated in Fig.~\ref{fig:capacity}. As we
clearly see, the ground state quickly approaches the Shannon's bound of
Eq.~(\ref{eq:shannon}) as $K$ increases. Thus, this particular CSP is very
promising from the point of view of lossy data compression. In
Fig.~\ref{fig:phasediagram} we show the phase diagram for $K=6$. The
\emph{static} energy is the ground state energy per variable also plotted in
Fig.~\ref{fig:capacity}; from the algorithmic point of view, this is the
performanece of the best possible algorithm which minimizes the number of
violated constraints. The \emph{dynamical} energy marks the appearance of a
regime where solutions group in many different well separated clusters (that
is, in order to go from one cluster to another one we should flip an
extensive number of variables); any local algorithm, as for example WP, will
be trapped at this dynamical threshold. The dynamical energy computed here
in the 1RSB approximation is believed to be an upper bound to the exact
one. Finally, the stability curve~\cite{montanariricci} indicates the range
of validity of the 1RSB formalism used to determine this phase diagram (in
particular, the ground state energy which we compute should be the exact one
for $\alpha<\alpha_{1rsb}$).

\begin{figure}
  \begin{center}
    \includegraphics[angle=-90,width=0.5\textwidth]{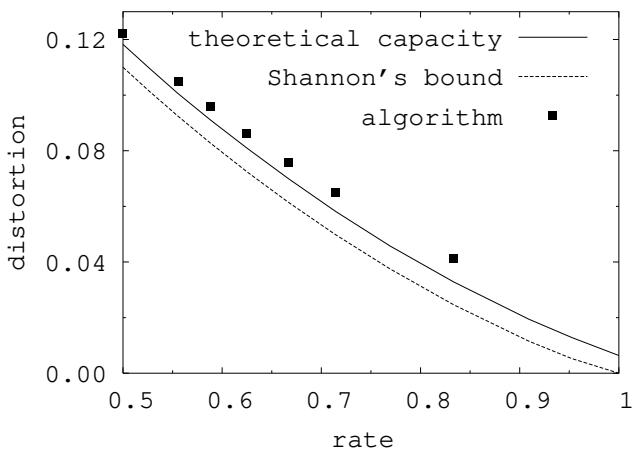}
  \end{center}
  \caption{The performance of the algorithm is plotted versus the rate
  $R=1/\alpha$ of the compression (here $K=6$, $N=1000$), together with the
  theoretical capacity and Shannon's bound.}
\label{fig:perf}
\end{figure}

So the theoretical properties of the random-node CSP are quite similar to
the parity check CSP (the XORSAT problem discussed in~\cite{merize}).  The
good point with respect to the parity check CSP is that in this case the SID
algorithm does converge in the unsat phase in a time which scales as $2^K N
\log N $, and gives very low energy states, {\it i.e.} nearly optimal global
configurations. In Fig.~\ref{fig:perf} we show the performance of the
compressor based on random gates, for $K=6$. The distortion achieved in
practice by the algorithm with $N=1000$ is close to the theoretical
capacity, which brings it a few \% above Shannon's bound. As shown in
Fig.~\ref{fig:capacity}, we expect the performance to improve with
increasing $K$ (at the price of an increase in computer time).

\emph{Conclusions.} We have shown, by using techniques borrowed from the
statistical physics of disordered systems, how one can use CSPs as a tool
for compressing data. In particular, the algorithmic performance of the
random gates -- CSPs based on non-linear nodes -- is found to be nearly
optimal, since the Shannon bound is reached at large $K$. The generalization
of the present approach to compression of data from a larger alphabet
(beyond binary input) looks like an interesting perspective.

\emph{Acknowledgments.} We thank D. Saad, M. Wainwright and J.S. Yedidia for
discussions.  S.~C. is supported by EC through the network MTR 2002-00307,
DYGLAGEMEM.  This work has been supported in part by the EC through the
network MTR 2002-00319 STIPCO and the FP6 IST consortium EVERGROW.


\end{document}